%Paper: hep-th/9408055
%From: Philip Tuckey <pht@mppmu.mpg.de>
%Date: Tue, 9 Aug 1994 15:33:11 +0200

%%%%% Plain TeX %%%%% Plain TeX %%%%% Plain TeX %%%%% Plain TeX %%%%%

% GEOMETRY AND DYNAMICS WITH TIME-DEPENDENT CONSTRAINTS
% by Jonathan M. Evans and Philip A. Tuckey
% for the proceedings of the Geometry of Constrained Dynamical Systems
% workshop, 15-18 June 1994, Newton Institute.

%PREPRINT VERSION

% MATH MACROS

\font\suff=cmmi6

\font\bbb=msbm10%msym10%cmr10

\def\del{\partial}

\def\half{ {\textstyle{1 \over 2}} }

\def\xt{ {\tilde x} }
\def\pt{ {\tilde p} }

\def\vpb{ {\bar \varphi} }
\def\vpt{\varphi_t}

\def\A{ {\hbox{\suff A} } }

\def\M{ {\hbox{\suff M} } }
\def\N{ {\hbox{\suff N} } }
\def\I{ {\hbox{\suff I} } }

\def\R{ {\hbox{\bbb R} } }

\def\G{ \Gamma}
\def\Gp{ \Gamma^* }
\def\Ge{ \bar \Gamma }
\def\Gpe{ \bar \Gamma^* }

\def\Xt{X_t}
\def\Xb{\bar X}

\def\z{ z^\M }
\def\v*{ {v^*} }
\def\os{ \omega^* }

\def\d{ {\rm d} }

%%%% New
\def\etal{{\it et al\/}}
\def\xd{\dot x}
\def\ca{}
\def\cb{}
\def\cas{} %\ca squared
 %\cb squared
%\def\P{\hbox{\cal P}}
%\def\P{P}
\def\halfp{ {\textstyle{1 \over 2\pi}} }

% END MACROS

\magnification=\magstep1

\line{\hfil CERN-TH.7392/94}
\line{\hfil MPI-PhT/94--45}
\rightline{hep-th/9408055}
\vskip 50pt

\centerline{\bf GEOMETRY AND DYNAMICS WITH TIME-DEPENDENT CONSTRAINTS}
\vskip 40pt
\centerline{
Jonathan M.~Evans${}^1$\footnote{*}{Supported by a fellowship
{}from the EU Human Capital and Mobility programme.}
and Philip A.~Tuckey${}^2$\footnote{**}{Supported by a fellowship
{}from the Alexander von Humboldt Foundation.}
}
\vskip 10pt
\centerline{ ${}^1$ Theoretical Physics Division, CERN}
\centerline{ CH-1211 Geneva 23, Switzerland}
\centerline{ e-mail: evansjm@surya11.cern.ch}
\vskip 5pt
\centerline{${}^2$ Max-Planck-Institut f\"ur Physik}
\centerline{Werner-Heisenberg-Institut}
\centerline{F\"ohringer Ring 6}
\centerline{D-80805 Munich, Germany}
\centerline{ e-mail: pht@mppmu.mpg.de}
\vskip 30pt
\centerline{\bf Abstract}
\vskip 5pt

\noindent
We describe how geometrical methods can be applied
to a system with explicitly time-dependent second-class constraints
so as to cast it in Hamiltonian form on its physical phase space.
Examples of particular interest are systems which require
time-dependent gauge fixing conditions in order to reduce them to their
physical degrees of freedom.
To illustrate our results we discuss the gauge-fixing of relativistic
particles and strings moving in arbitrary background electromagnetic
and antisymmetric tensor fields.

\vskip 30pt

\centerline{\sl To appear in the Proceedings of the Conference on
Geometry of Constrained Dynamical}
\centerline{\sl Systems, Isaac Newton Institute for Mathematical Sciences,
Cambridge, U.K.,}
\centerline{\sl 15th-18th June 1994, ed J.M.~Charap (CUP)}
\vfil

\line{CERN-TH.7392/94 \hfil}
\line{MPI-PhT/94--45 \hfil}
\line{August 1994 \hfil}
\eject

%%%% THE BUSINESS

\noindent
{\bf 1. Introduction}
\vskip 5pt

\noindent
A Hamiltonian dynamical system can be described geometrically by a
phase space manifold $\G$ (of dimension $2d$ say)
equipped with a symplectic form $\omega$ and
a Hamiltonian function $H$ (see Abraham and Marsden (1978) and Arnol'd (1978)).
The condition that $\omega$ is symplectic means that it is a non-degenerate
closed two-form, so it can be used to introduce a Poisson bracket
$\{ \, \hbox{,} \, \}$ on $\G$.
The evolution of the system in time $t$ is given by a particular set
of trajectories on $\G$, parametrized by $t$, such that
Hamilton's equation
$$
{\d f \over \d t} = {\del f \over \del t} + \{ f , H \}
\eqno(1) $$
holds for any time-dependent function $f$ on $\G$.

Since the seminal work of Dirac (1950,1958,1964) there has been
intensive study of
systems of this type which can be consistently constrained to some
physical phase space manifold $\Gp$ (of dimension $2n$ say) which is
embedded in $\G$ in a manner we now describe.
In the most general case the embedding of $\Gp$ in $\G$ can depend on time and
it must therefore be defined by a family of maps
$$
\vpt : \Gp \to \G \ ,
\eqno (2)$$
depending smoothly on $t$, each of which is a diffeomorphism onto its
image $X_t \subset \G$.
We assume that each of the trajectories on $\G$ for which (1) holds
has the property that it always lies in the subspaces
$X_t$ for each $t$, or else that it always lies in the complements of
these spaces.
It is clear that trajectories of the former type correspond
exactly under the embedding (2) to trajectories on $\Gp$,
and one can attempt to reformulate the dynamics for this subclass of
trajectories in a manner which is intrinsic to $\Gp$.

We define $\os$ on $\Gp$ at time $t$ by pulling back $\omega$ using
$\vpt$ and we assume that this is also a symplectic form (albeit a
time-dependent one in general). We can then use $\os$ to introduce the Dirac
bracket $\{ \, \hbox{,} \, \}^*$ on $\Gp$. The key issue which we shall
address here is whether, for a given choice of embeddings (2), one can find
a Hamiltonian function $H^*$ such that Hamilton's equation
$$
{\d f \over \d t} = {\del f \over \del t} + \{ f , H^* \}^*
\eqno(3)$$
holds for any time-dependent function $f$ on $\Gp$.
When the embeddings (2) are independent of time, (3) follows easily
{}from (1) with $H^* = H$. In the general case, however, the dynamics on
$\Gp$ is determined not just by the dynamics on $\G$ but also by the
time-dependence of the embeddings $\vpt$, and under these circumstances
it is non-trivial to determine whether (3) holds for some function
$H^*$.

The most compelling reason for studying this general situation is the fact
that gauge choices with explicit time dependence are essential in
order to restrict systems which are invariant under
time-reparametrizations, such as the relativistic particle, string
or general relativity, to their physical degrees of freedom
(but see Henneaux {\it et al\/} (1992) for possible modifications of
the action to allow other gauge choices).
Here we summarize the solution of this problem given in Evans and
Tuckey (1993) and we clarify some related issues.
(We have recently learned that Mukunda (1980) has previously obtained
results which are locally equivalent to ours using an algebraic approach.
Related work from
the Lagrangian point of view appears in (Ra\~nada 1994).)
We then give some new examples, extending the treatment of the relativistic
particle in a background field (Evans 1993) to the case of a string
in an arbitrary antisymmetric tensor background.
\vskip 10pt

\noindent
{\bf 2. Extended phase space and constrained dynamics}
\vskip 5pt

\noindent
We define extended phase space to be $\Ge = \G \times \R$, where the second
factor is time. We can, in a natural way, regard $H$ and $\omega$ as
living on $\Ge$ (by pulling back using the projection map) and we
define the contact form on $\Ge$ to be
$$
\Omega = \omega + \d H \wedge \d t \ .
\eqno (4) $$
(In Evans and Tuckey (1993) $\Omega$ was called the Poincar\'e-Cartan
two-form; in Abraham and Marsden (1978) $\Omega$ is introduced as an
example of a contact structure.)
Any trajectory on $\G$ parametrized by $t$ is clearly equivalent to a
trajectory on $\Ge$ with parameter $s$ chosen such that $\d t/\d s$ is
nowhere zero. Let $V$ be the tangent vector to the
trajectory on $\Ge$. Then Hamilton's equation (1) is precisely the condition
$$
i(V) \Omega = 0
\eqno(5) $$
(where $i(V)$ denotes interior multiplication of a form
by the vector field $V$).

When the system is constrained we can similarly define extended
physical phase space to be $\Gpe = \Gp {\times} \R$.
The family of embeddings (2) is equivalent to the single embedding
$$
\vpb : \Gpe \rightarrow \Ge \ , \qquad \vpb (x,t) = (\vpt(x) ,t) \ ,
\eqno(6) $$
which is a diffeomorphism onto its image
$\Xb = \{ (x,t) : x \in \Xt, t \in \R \} \subset \Ge$
(assuming, as stated earlier, that $\vpt$ varies smoothly with $t$).
Define the form $\Omega^*$ on $\Gpe$ to be the pull back of $\Omega$
using $\vpb$.
In general this has the structure
$$
\Omega^* = \omega^* + (\d H + Y) \wedge \d t
\eqno (7) $$
for some one-form $Y$. (Here we use the fact that any time-dependent
form on $\Gp$ can be regarded as a smooth form on
$\Gpe$; when the form is time-independent this reduces to pulling
back using the projection map.)

Any solution of Hamilton's equation (1) which lies in
$\Xb$ clearly corresponds to a trajectory
in $\Gpe$ with tangent vector $V^*$ which satisfies
$$ i(V^*) \, \Omega^* = 0 \ .
\eqno (8) $$
By comparison with (4) and (5) we see that
Hamilton's equation (3) holds on $\Gp$ if and only if
$$
Y = \d K \ {\rm mod}\ \d t \qquad \hbox{\rm and then}
\qquad H^* = H + K
\eqno(9)$$
for some function $K$ on $\Gpe$.
One can show that this holds locally ({\it ie.\/} in any
contractible region on $\Gp$) if and only if $\os$ is independent of
time on $\Gp$, a fact which will prove useful later.

To discuss specific examples it is convenient to introduce
on $\Gamma$ local coordinates $\z$, $M = 1 , \ldots , 2d$.
The subsets $X_t \subset \G$ are defined by
a set of time-dependent constraint functions
$\psi^\I (\z , t)$, $I = 1, \ldots , 2(d{-}n)$, which are, in the
language of Dirac (1950,1958,1964), second-class.
The fact that these constraint functions are preserved in time is
equivalent to our initial assumption that there exists a subset of
trajectories confined to the subspaces $X_t$.
The condition that the constraint functions are second-class
is equivalent to our assumption that the form $\os$ is symplectic.

If $\xi^\A$, $A = 1, \ldots , 2n$, are local coordinates on
$\Gp$ then the embeddings $\vpt$ or $\vpb$ allow us to regard the $z^\M$
as time-dependent functions of these variables on $\bar X$,
and we have the explicit expression
$$
Y = - \, {\del z^\M \over \del t} \,
{\del z^\N \over \del \xi^\A} \,
\omega_{\M \N} \, \d \xi^\A
= - \, \omega_{\M \N} \, {\del z^\M \over \del t} \, \d z^\N
\quad{\rm mod}\ \d t
\eqno (10) $$
for the one-form appearing in (9).
It is convenient in practice to specify $\vpt$ or $\vpb$ by
giving explicit expressions for a set of functions $\xi^\A(\z,t)$, which
we call physical variables; on restriction to $\bar X$ these
functions define (the inverses of) these embeddings in terms of the local
coordinates.
\vskip 10pt

\noindent
{\bf 3. Remarks}
\vskip 5pt

\noindent
Our result (9)
establishes necessary and sufficient conditions for a
family of embeddings $\vpt$, or a choice of physical variables $\xi^\A
(z^\M , t)$, to result in a Hamiltonian time-evolution equation (3) on
$\Gp$.
For a given set of constraint subspaces $X_t$, or equivalently a set
of constraint functions $\psi^\I (z^\M , t)$, a family of embeddings or
physical variables having this property always exists locally.
This follows from Darboux's Theorem, which tells us that locally we
can find embeddings $\vpt$ or choose coordinates $( \xi^\A ) = (q^\alpha ,
p_\alpha)$ on $\Gp$ such that $\os = \d q^\alpha \wedge \d p_\alpha$.
Since this
expression is manifestly independent of time on $\Gp$, it satisfies
the criterion which we gave following (9). On the other hand, there
are clearly many embeddings or choices of physical variables for which
(3) will not hold, as can be seen by performing an arbitrary
time-dependent coordinate transformation to make $\os$ time-dependent.

Our result does not tell us how to explicitly construct a set of embeddings
or physical variables with the required property, and in general this
remains an open problem. A partial result in this direction is case (B) of
Evans (1991). This applies to a system with time-independent gauge symmetry
generators which has imposed on it a set of time-dependent gauge-fixing
conditions involving some subset of canonical variables which all commute
under the Poisson bracket. It is worth pointing out that if these canonical
variables are regarded as configuration space coordinates in some
equivalent Lagrangian description, then the result in question can also be
obtained by first gauge-fixing the Lagrangian and then passing to the
Hamiltonian formalism.
The new examples we shall present below lie outside the scope of case (B)
of Evans (1991). Thus the result (9) is still useful for finding good sets of
physical variables, even though it offers no general method for doing so.

Finally we emphasize that our main motivation for the work summarized here
is the reduced phase space approach to the canonical quantisation of
systems which require time-dependent gauge choices.   Even at the classical
level, a system whose time evolution is not described by an equation of the
form of (3) falls outside the realm of conventional Hamiltonian mechanics.
In passing to the quantum theory, (3) becomes the Heisenberg equation of
motion, which guarantees the existence of a unitary time evolution
operator. In the absence of a classical evolution equation of the form of
(3), Gitman and Tyutin (1990a) have given an alternative prescription for
the Heisenberg quantum evolution equation, in which extra terms appearing
on the right hand side of (3) are taken over. This approach is complicated
by the difficulty in obtaining an explicitly unitary time evolution. The
consideration of simple examples such as the relativistic particle in an
arbitrary background electromagnetic field (Evans 1993) reveals that our
approach can be much simpler -- compare with Gitman and Tyutin (1990b),
Gavrilov and Gitman (1993). Batalin and Lyakovich (1991) have also
considered the quantization of systems with time-dependent Hamiltonian and
constraints.
\vskip 10pt

\noindent
{\bf 4. Examples}
\vskip 5pt

\noindent
We shall now apply our result (9) to discuss the gauge-fixing of relativistic
particles and strings moving in $d$-dimensional Minkowski space-time with
background gauge fields. We shall take coordinates $x^\mu$
on Minkowski space-time which are either `orthonormal'
with $\mu = 0, \ldots , d{-}1$, or
of `light-cone' type with
$\mu = +, -, 1, \ldots , d{-}2$, so that the flat metric has components
$- g_{00} = g_{11} = \ldots = g_{d{-}1 d{-}1} =
g_{+-} = g_{-+} = 1$ and all others vanishing.
(This means that $x^\pm = (x^{d-1} \pm x^0)/\sqrt2$ agreeing with the
conventions of Evans (1993) but not Evans (1991).)
It is useful to set up some conventions regarding indices which
will allow us to deal with temporal and light-cone gauge conditions in a
uniform way.
Thus we shall let the single index $n$ on any vector denote either $0$ or $+$,
and we shall label the remaining components by
$a = 1, \ldots , d{-}1$ or $a = -, 1, \ldots, d{-}2 $ respectively.
We shall also find it useful to denote the `transverse' components by
$i=1, \ldots, d{-}2$.
In what follows the ranges of these indices will always be understood.
\vskip 5pt

\noindent
{\bf 4.1 Relativistic particle in an electromagnetic field}

\noindent
A particle of mass $m$ and charge $e$ moving in an arbitrary
electromagnetic field $A_\mu(x^\nu)$ can be described by the
Lagrangian
$$
L = - \left[\,m\sqrt{-\xd^2} + e A_\mu(x^\nu)\,
\xd^\mu\right]\ .
\eqno(11)$$
Here $x^\mu(t)$ is the particle's trajectory,
$t$ is a parameter along the worldline, and $\xd=\d x/\d t$.
Introducing the canonical momentum $p_\mu = \del L / \del \xd^\mu$
conjugate to $x^\mu$, we have coordinates $(x^\mu,p_\mu)$ on phase space,
with Poisson bracket $\{x^\mu,p_\nu\} = \delta^\mu_\nu$. There is a single,
first-class constraint
$$
\phi =
(p+eA)^2 + m^2 = 0 \ .
\eqno(12)$$
The Hamiltonian is $H=\lambda\phi$, where $\lambda$ is an arbitrary
(time-dependent) function on phase space.

Consider the class of gauge-fixing conditions of the form
$$
x^n = f(p_\mu,t) \ ,
\eqno(13)$$
where $f$ can be any function of its arguments which defines a good
gauge choice (we shall make no attempt to be more precise concerning this last
point). We define a set of physical variables
$$
(\xi^\A) = (x^{a*},p_a) \quad{\rm where}\quad
x^{a*} = x^a - \int dp_n {\del f\over\del p_a}
\eqno(14)$$
(partial derivatives and integrals of $f$ are to be understood in terms
of the functional dependence given by (13))
and it is clear that in principle
the equations (12), (13) and (14) allow us to express all quantities
as functions of $(\xi^\A , t)$.
We claim that the system can then be
described by these physical variables together with a Hamiltonian
$$
H^* = -\int dp_n \, {\del f\over\del t} \ ,
\eqno(15)$$
which is valid for any background gauge field $A_\mu$ and any function $f$.

The explicit restriction to physical phase space is of course very involved
for a general background field.
In principle we can substitute from (13) and (14) into (12) to find
$p_n$ as a function of the physical variables and time, and substitution
of this result back into (13) and (14) then determines all the $x^\mu$
as functions of $(\xi^\A , t)$.
Fortunately, it is not necessary to carry out this elimination explicitly
in order to verify that our chosen physical variables do indeed satisfy
the criterion (9) leading to the general expression for the Hamiltonian
given above. This is because
the explicit time dependence of $x^\mu$ enters only through $p_n$ and
$f$; and by using this fact it is easy to calculate from (10)
that $Y = -\d(\int d p_n\,\del f/\del t) \ {\rm mod}\ \d t$. Since
the original Hamiltonian $H$ vanishes when $\phi=0$, the result follows.

Examples are:
$$
%\eqalign{
%&
x^n = t \quad{\rm giving}\quad x^{a*} = x^a \, ,
\quad H^* = -p_n \ ,
%\cr
%&x^+ \! = t \quad{\rm giving}\quad x^{a*} = x^a
%\quad{\rm and}\quad H^* = -p_+ \ , \cr
%}
\eqno(16)$$
which reproduces the temporal and light-cone results of Evans (1993);
$$\eqalign{
&x^+ \! = p^+ t \quad{\rm giving}\quad x^{-*} = x^- \! - p^- t \, , \quad
x^{i*} = x^i \, , \quad H^* = - p_+ p_- \ ;\cr
&x^0 = p^0 t \, \quad{\rm giving}\quad x^{a*} = x^a \, ,
\quad H^* = \half p_0^2 \, . \cr
}\eqno(17)$$
These expressions are deceptively simple in appearance
because they represent very
complicated functions of the physical variables in the case of a general
background.
It is interesting that the Hamiltonians have universal forms in
terms of the original momenta, in the sense that the dependence on the
background field enters only through these particular functions.
\vskip 5pt

\noindent
{\bf 4.2 Relativistic closed string in an antisymmetric tensor field}

\noindent
A closed string moving in an arbitrary background antisymmetric tensor
field $B_{\mu\nu}(x^\rho)$
can be described by the Lagrangian
$$
L = - \int_0^{2\pi}d\sigma\left[
\ca\left(\left(\xd . x'\right)^2 - \left(\xd\right)^2
\left(x'\right)^2\right)^{1/2} +
\cb B_{\mu\nu}(x^\rho)\,\xd^\mu{x'}^\nu \right] \ .
\eqno(18)$$
Here $x^\mu(t,\sigma)$ describes the string's trajectory,
$t$ and $0\leq\sigma\leq  2\pi$ parametrize the world-sheet,
and $\xd = \del x/\del t$, $x' = \del x/\del\sigma$.
Introducing the momentum
$p_\mu(\sigma) = \delta L/\delta \xd^\mu(\sigma)$
conjugate to $x^\mu(\sigma)$ as usual, we have
coordinates $(x^\mu(\sigma),p_\mu(\sigma))$ on phase
space, with Poisson bracket $\{x^\mu(\sigma),p_\nu(\sigma')\} =
\delta^\mu_\nu\delta(\sigma-\sigma')$. The only constraints are first-class
and are given by
$$
\left(p_\mu+\cb B_{\mu\nu}x'^\nu\right)^2
+ \cas \left(x'\right)^2 = 0 \ ,
\eqno(19)$$
$$
x' . \, p = 0 \ .
\eqno(20)$$
Again the Hamiltonian $H$ is proportional to the constraints.

It is useful to introduce the position zero-mode and total momentum of the
string by
$$
X^\mu(t) = {1\over2\pi}\int_0^{2\pi}d\sigma\,x^\mu(t,\sigma)\ , \qquad
P_\mu(t) = \int_0^{2\pi}d\sigma \, p_\mu(t,\sigma) \ .
\eqno(21)$$
The factor of $2\pi$ ensures $X^\mu$ and $P_\mu$ are
conjugate variables.
One can then write a decomposition of the string fields
$$
x^\mu = X^\mu + \xt^\mu
\, , \qquad
p_\mu = \halfp P_\mu + \pt_\mu \, ,
\eqno(22) $$
where the tilded variables represent the oscillator degrees of freedom.

We consider the class of gauge-fixing conditions of the form
$$
x^n = f(P_\mu,t) \ ,
\eqno(23)$$
$$
p^n = \halfp P^n \ ,
\eqno(24)$$
where $f$ is, as before, any
function of its arguments which provides a good gauge-fixing condition.
The gauge conditions and constraints allow for the complete elimination of
one pair of string position and momentum variables corresponding to the
direction in space-time labelled by $n$, and they allow
also for the elimination of one additional set of string oscillators.
We introduce a set of physical variables
$$
(\xi^\A) = (X^{a*},P_{a}, \xt^{i}(\sigma),\pt_i(\sigma))
\quad{\rm where}\quad
X^{a *} = X^{a } - \int dP_n {\del f\over\del P_{a}}
\eqno(25)$$
(comments similar to those following (14) apply)
and with a little thought one can see that the equations (19), (20), (23),
(24) indeed allow us in principle to express all the original variables in
terms of $(\xi^\A,t)$.
At this stage the analysis looks very like that for the particle, at least
as far as the string zero mode and total momentum variables are concerned.
For the case of a light-cone gauge ($n = +$) this comparison is accurate:
for the physical variables given above one can again calculate $Y$ from
(10) and deduce that (9) holds, yielding a Hamiltonian
$$
H^* = -\int d P_n \, {\del f\over\del t} \ ,
\eqno(26)$$
which is valid for any background field $B_{\mu \nu}$ and any gauge-fixing
function $f$.
For the case of a temporal gauge condition ($n = 0$), however, the physical
variables written above do not satisfy (9) in general, unless the background
has some special symmetry. The previous arguments break down because
the solutions for the redundant oscillator variables $\xt^{d-1}$ and
$\pt_{d-1}$ in terms of $(\xi^\A , t)$ can depend explicitly on time for a
general background field.
This difficulty is absent if, for example, the background vanishes,
$B_{\mu \nu} = 0$, and then the physical variables and Hamiltonian written
above hold for any function $f$.

Examples are:
$$
x^n = t\quad{\rm giving}\quad X^{a*} = X^a\ ,
\quad H^* = -P_n \ ;
\eqno(27)$$
and
$$\eqalign{
x^+ \! = P^+ t \!&\quad{\rm giving}\quad
X^{- *} = X^{-} \! - P^- t \, , \ \
X^{i*} = X^i \ ,
\quad
H^* = - P_+ P_- \ ; \cr
x^0 = P^0 t &\quad{\rm giving}\quad X^{a*} = X^a \ ,
\quad
H^* = \half P_0^2 \ . \cr
}\eqno(28)$$
These last examples generalise previous light-cone and temporal gauge-fixing
constructions (Goddard \etal\ 1973, Scherk 1975, Goddard \etal\ 1975).
\vskip 10pt

\noindent
{\bf Acknowledgments}
\vskip 5pt

\noindent
We would like to thank Prof.~J.M.~Charap for organising this very
stimulating conference.
\vskip 10pt

% REFERENCES

\def\npb{{Nucl. Phys.}}
\def\plt{{Phys. Lett.}}

\def\cjm{{Canad. J. Math.}}

\def\cqg{{Class. Quantum Grav.}}
\def\imp{{Int. J. Mod. Phys.}}

\def\prs{{Proc. Roy. Soc.}}

\def\rmp{{Rev. Mod. Phys.}}
\def\psc{{Phys. Script.}}
\def\jmp{{J. Math. Phys.}}

\noindent
{\bf References}
\vskip 5pt

\noindent
Abraham R and Marsden J E, 1978, {\it Foundations of mechanics}
Benjamin/Cummings, 1978
\vskip 5pt

\noindent
Arnol'd V I, 1978, {\it Mathematical methods of classical mechanics}
Springer, 1978
\vskip 5pt

\noindent
Batalin I A and Lyakovich S L, 1991, in {\it Group theoretical
methods in physics} Nova Science, 1991, vol.~2, p.~57
\vskip 5pt

\noindent
Dirac P A M, 1950, \cjm\ {\bf 2} 129
\vskip 5pt

\noindent
Dirac P A M, 1958, \prs\ {\bf A246} 326
\vskip 5pt

\noindent
Dirac P A M, 1964, {\it Lectures on quantum mechanics} Academic, 1964
\vskip 5pt

\noindent
Evans J M, 1991, \plt\ {\bf B256} 245
\vskip 5pt

\noindent
Evans J M, 1993, \cqg\ {\bf 10} L221
\vskip 5pt

\noindent
Evans J M and Tuckey P A, 1993, \imp\ {\bf A8} 4055
\vskip 5pt

\noindent
Gavrilov S P and Gitman D M, 1993, \cqg\ {\bf 10} 57
\vskip 5pt

\noindent
Gitman D M and Tyutin I V, 1990a, {\it Quantization of fields with
constraints} Springer, 1990
\vskip 5pt

\noindent
Gitman D M and Tyutin I V, 1990b, \cqg\ {\bf 7} 2131
\vskip 5pt

\noindent
Goddard P, Goldstone J, Rebbi C and Thorn C B, 1973, \npb\ {\bf B56} 109
\vskip 5pt

\noindent
Goddard P, Hanson A J and Ponzano G, 1975, \npb\ {\bf B89} 76
\vskip 5pt

\noindent
Henneaux M, Teitelboim C and Vergara J, 1992, \npb\ {\bf B387} 391
\vskip 5pt

\noindent
Mukunda N, 1980, \psc\ {\bf 21} 801
\vskip 5pt

\noindent
Ra\~nada M F, 1994, \jmp\ {\bf 35} 748
\vskip 5pt

\noindent
Scherk J, 1975, \rmp\ {\bf 47} 123
\bye